\documentstyle[aps,preprint]{revtex}
\def\lan{\langle}
\def\ran{\rangle}
\begin{document}
\draft
\title{ Effect of nonlinearity on the dynamics of a particle 
in dc field-induced systems}
\author{P. K. Datta$^{\dag}$ \cite{mail1} and A. M. Jayannavar$^{\ddag}$ \cite{mail2}}
\address{$^{\dag}$Mehta Research Institute of Mathematics and Mathematical
Physics, Allahabad- 211 019, India\\
$^{\ddag}$Institute of Physics, Bhubaneswar-751 005, India}
\maketitle
\begin{abstract}
Dynamics of a particle in a perfect chain with one nonlinear impurity
and in a perfect nonlinear chain under the action of dc field is studied
numerically. The nonlinearity
appears due to the coupling of the electronic motion to
optical oscillators which are treated in adiabatic approximation.
 We study for both the low and high values of field strength.
Three different range of nonlinearity is obtained where the dynamics
 is different. In low and intermediate range of
nonlinearity, it reduces the localization. In fact in the
intermediate range subdiffusive behavior in the perfect nonlinear
chain is obtained for a long time. In all the cases
a critical value of nonlinear strength exists where self-trapping transition
 takes  place. This critical value depends on the system and the 
field strength. Beyond the self-trapping transition nonlinearity enhances
 the localization.

\end{abstract} 
\narrowtext

The application of dc field on a charge particle in a perfect
periodic lattice would cause the oscillatory movement of the particle
with frequency $\omega = e E a/ \hbar$. Here $e$ is the charge of the
particle, $a$ is the lattice spacing and $E$ is the strength of the
electric field. This oscillation is called Bloch oscillation \cite{1}.
 The
rapid oscillation of the charge particle may act as a fast emitter of
electromagnetic radiation. However, the detection in bulk sample is
almost impossible due to the fact that scattering time is much
smaller than the period of Bloch oscillator. Esaki and Tsu \cite{2} 
first
studied the Bloch oscillation in superlattice structure (SL) which
have long periods that makes the possible application for obtaining
electromagnetic radiation in terahertz range. Recently these phenomena
have been confirmed by various laboratory experiments \cite{3}. A simple
theoretical approach which captures the underlying physics is to use
tight-binding models.
Recently, the dynamical localization of a carrier has been studied in
a one-dimensional perfect lattice with one linear impurity \cite{4}. The
resonance occurs for a particular value of the applied field and the
strength of the impurity. This is observed only when the electric field is
very high \cite{14}. In this case the two on-site energies
coincide which enhances the hopping of the carriers between the two
degenerate sites. On the other hand, for weak field the degenerate sites are
far apart from each other and as a result there is almost complete
localization at the impurity site.

In this work we numerically study the dynamics of a charged particle in
a linear
chain with a single nonlinear impurity and in a perfect nonlinear
chain under the action of dc electric field. The
general equation of motion for the system is given as
\begin{equation}
i \hbar\frac{dc_m}{dt} = V (c_{m+1} + c_{m-1}) - (\epsilon_m + \chi_m |c_m|^2
 + e a m E) c_m. \label{sch}
\end{equation}
Here $c_m(t)$ is the probability amplitude of the particle at
site $m$ at time $t$, $V$ is the nearest-neighbor transfer matrix
element, $\epsilon_m$ and $\chi_m$ are on-site energy and
nonlinearity strength of the $m$-th site, respectively. Here, we
assume $e = a = \hbar = V = 1$. The nonlinearity arises due to the
coupling of quasiparticles with optical oscillators in adiabatic
approximation \cite{7}. When the electric field $E$ is zero the
equation is called one-dimensional discrete nonlinear
Schr\"odinger equation (DNLS) \cite{7,9,10,11,13}. The
interesting property of DNLS equation is that self-trapping transition
occurs if the nonlinearity strength exceeds a critical value. The
mean-square displacement (MSD) for perfect nonlinear systems \cite{9} as
well as  random 
nonlinear systems \cite{10,11} goes as $\sim t^2$.
It has been found that the time-averaged site-probability is a good
candidate to study the dynamics of the system \cite{11,13}. So, we
mainly study here the time-averaged probability of different sites of
the systems discussed above for low as well as high electric field.
The time-averaged probability at $m$-th site is defined as
\begin{equation}
\lan P_m \ran = \lim_{T \rightarrow \infty}
\frac{1}{T}\int_0^T|c_m(t)|^2 \label{tavp}
\end{equation}
with \( |c_m(0)|^2 = \delta_{m,0}\). We solve the coupled nonlinear
differential equations numerically using 4-th order Runge-Kutte method.
To avoid the boundary effect we use self-expanding lattice \cite{5}. For
 time-averaging we have taken $T = 2000$. The accuracy of numerical 
investigation is checked through the total probability.
To examine the long time behavior of the system we also study the
particle propagation and MSD. The MSD is defined as
\begin{equation}
< m^2 > = \sum_{m = -\infty}^{\infty} m^2 |c_m(t)|^2 \label{mean}
\end{equation}
with the same initial as above.

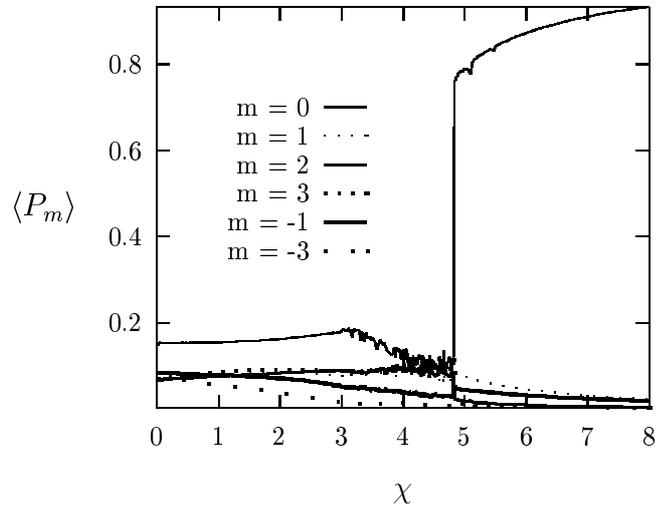
\begin{figure}
\caption{Plot of time-averaged probability $< P_m >$ as a function 
of $\chi$ for different values of $m$ in a perfect lattice containing
a nonlinear impurity. The value of $E$ is $0.5$.}
\label{slp}
\end{figure}

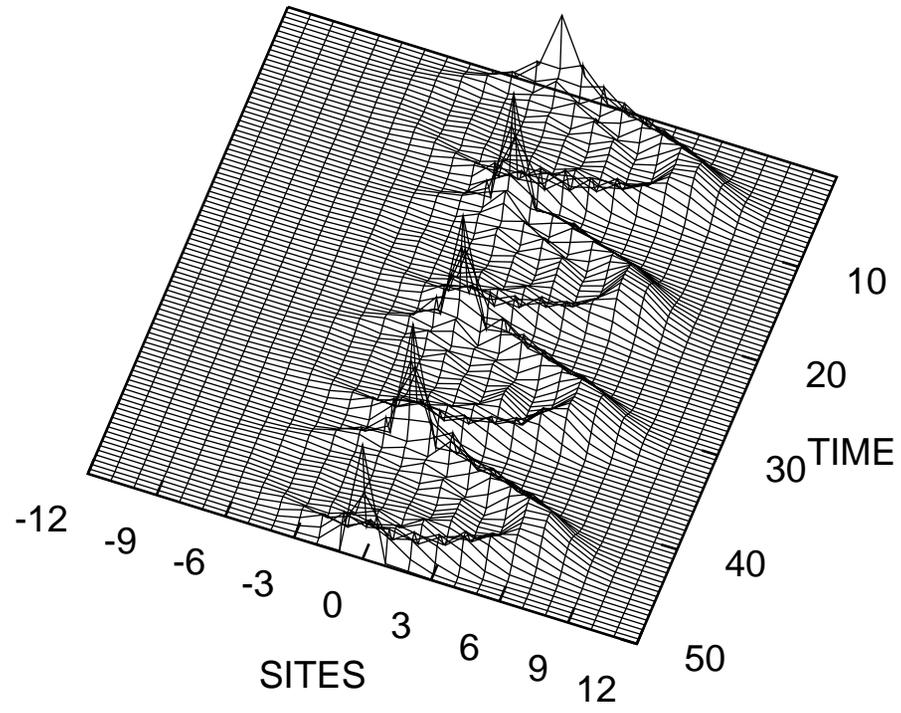
\begin{figure}
\caption{Electronic probability propagation profile
 as a function of time $(t)$ for $\chi = 2.5$ and $E = 0.5$ in the same
 system as in Fig. \ref{slp}.}
\label{prop}
\end{figure}

We first study the dynamical properties of a dc field-induced
perfect chain containing only one nonlinear impurity at the zeroth
site. The equation of motion is same as Eq. \ref{sch} with $\epsilon_m = 0$
 for all $m$
and $\chi_m = \chi \delta_{m,0}$. The nonlinearity alters the effective
strength of the zeroth site dynamically as it depends on the occupation
 probability at that site. In a single linear impurity problem the
degeneracy between the zeroth site and $m$th site occurs when $\epsilon_0 =
m E $ (where $\epsilon_0$ is the defect strength at zeroth site)
\cite{4,14}. But
here, the
degeneracy ($\chi |c_0(t)|^2 = m E$) may appear
 dynamically i.e. the
strength of the zeroth site and any other site become equal at different
 instants of time. This is due to the oscillatory behavior of 
$|c_0(t)|^2$ with time $t$. For low field (e.g. $E = 0.5$) the time-averaged
probability of a few sites as a
function of nonlinearity strength $\chi$ is shown in Fig. \ref{slp}.
 In this study we can define three regions of $\chi$ where the
 behavior changes significantly: (1) for small values of nonlinearity
 strength $(\chi < 3)$ where we find smooth behavior of
 time-averaged probability of the sites under study; (2) intermediate
 values of $\chi, (3 < \chi < 4.8$) where we obtain small
 fluctuations and (3) high values of $\chi (> 4.8)$ where the
 particle is confined at the zeroth site with maximum probability.
 In region (1) for very
 small values of $\chi$ we do not find any significant
 change. However, if we increase the value of
 $\chi$ further particle gets confined towards the right side $(m > 0)$
of zeroth
 site with maximum probability (see Fig. \ref{prop}). Due
to the presence of nonlinearity at zeroth site, it becomes dynamically 
degenerate with the sites right to the zeroth site. On the other hand,
 the energy difference between the zeroth site and its left neighbor $(m = -1)$
 is in general greater than the applied electric field.
 Consequently, the
 particle prefers to move towards $m > 0$. The MSD oscillates and its
 amplitude also changes with time.
In region 2, the fluctuations occur due to the chaotic behavior
 of site-probabilities \cite{13}.
 It should be noted that the values of $\lan P_m \ran$
with $m \geq 0$ under study are almost equal. This implies particle localizes 
within a region 
uniformly. The region of localization
depends on the dynamical degeneracy of the furthest site with
 the zeroth site (i.e. the site which is degenerate with zeroth site at time
$t = 0$). This behavior is a sharp contrast to the linear impurity 
problem \cite{4}. In linear impurity problem the particle gets localized at the
 defect site more and more with increasing the defect strength. In the
 fluctuation regime the value of MSD is even larger than that of perfect
 field-induced systems. Thus, the nonlinearity reduces field-induced
localization. Beyond the fluctuations regime we find a 
self-trapping transition at $\chi_{cr} \sim 4.8$ where the maximum
probability is gathered at the initial excitation site (i.e. zeroth site). 
Just below 
$\chi_{cr}$, as $\chi$ increases the time spending by the particle at
 zeroth site with maximum probability also increases. This indicates the
 self-trapping transition at $\chi_{cr}$. However, after $\chi_{cr}$ if we
increase the value of $\chi$ the time-averaged probability of the zeroth site
increases and of other sites it decreases. The value of MSD decreases
even than that of field-induced perfect system. In the
absence of electric field the self-trapping transition occurs at
 $\chi_{cr} \sim 3.2$ \cite{11,13}. Thus, the value of $\chi_{cr}$ increases
 due to the presence of electric field. The particle has a 
tendency to move to the dynamical degenerate sites.
Consequently, to get the self-trapping transition larger value of $\chi$
is required.

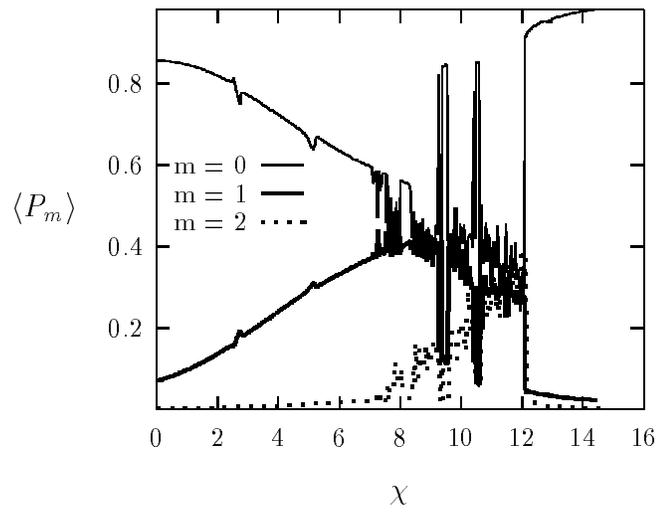
\begin{figure}
\caption{Same as Fig. \ref{slp} but $E = 5$.}
\label{shp}
\end{figure}

In Fig. \ref{shp} we have plotted the time-averaged probability of 
different sites as a function $\chi$ for high electric field (e.g. $E
= 5$). Here, $\lan P_0 \ran$ and $\lan P_1 \ran$
initially decreases and increases respectively with increasing $\chi$.
Beyond $\chi \sim 7.7$ we find the fluctuations in the time-averaged
probabilities. The self-trapping transition occurs at
$\chi_{cr} \sim 12.1$. Without any nonlinearity the particle gets localized
 at the zeroth site with maximum probability. As we increase the value
of nonlinearity the amplitude of 
the occupation probability of zeroth and its right neighbor site $(m = 1)$
increases. Thus the time-averaged probability of these two sites decreases
 and increases respectively with increasing $\chi$. In the fluctuation regime
 particle first gets localized within the sites
$m = 0$ and $1$ with maximum probability. Here, $|c_0(t)|^2 \leq 1$ and it has
oscillatory behavior. So, to obtain the degeneracy $(\chi |c_0(t)|^2 = E)$
between zeroth and its right neighbor site at many instants of
 time, the value
of $\chi$ should be larger than $E$. This is exactly obtained.
As we increase $\chi$ the site $m = 2$ also starts becoming
degenerate with zeroth site at different instants of time. Thus
 $\lan P_2 \ran$ gradually increases with increasing $\chi$
and at some values of $\chi$ we find $\lan P_0 \ran$,
 $\lan P_{1} \ran$ and $\lan P_{2} \ran$ are almost equal. This means
 that the particle gets localized within the three sites $m = 0, 1$ 
and $2$. To find the long time behavior we also study the particle
 propagation. The behavior is consistent with that of
 time-averaged probability. It should be noted that in single linear
impurity problem the resonance
occurs between the two sites \cite{4}. However, in the study of MSD we
find  the nonlinearity suppresses the field-induced localization as like low
field case. 
 Beyond $\chi_{cr}$ we find increasing of $\lan P_0 \ran$ and decreasing of
the time-averaged probability of all other sites with increasing
$\chi$. This means that the particle gets localized at the initial
excitation site with maximum probability. The study of MSD shows
the enhancement of the localization. It should be noted that the
critical value of 
$\chi$ for self-trapping transition increases with increasing the strength
of dc field. As we increase the field strength larger values of $\chi$ are
 required to get the degeneracy between the sites at many instants of time.
The particle prefers to move to the degenerate sites. Consequently, larger
 $\chi$ values are required to get the self-trapping transition.

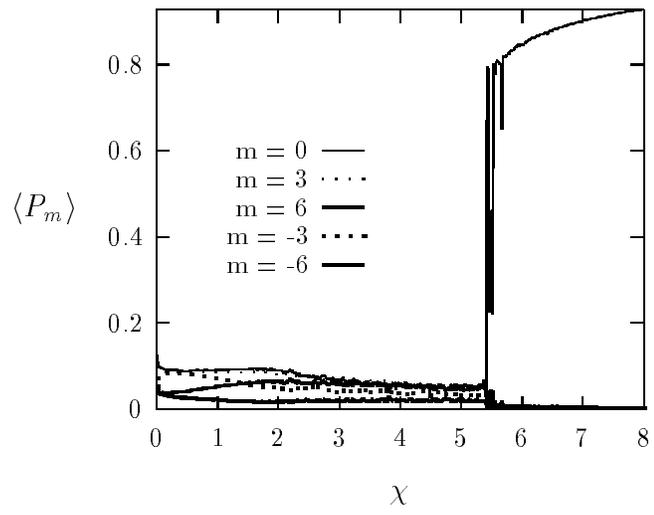
\begin{figure}
\caption{Time-averaged probability $< P_m >$ as a function $\chi$ for 
different values of $m$ in a perfect nonlinear chain. Here $E = 0.5$.}
\label{plp}
\end{figure}

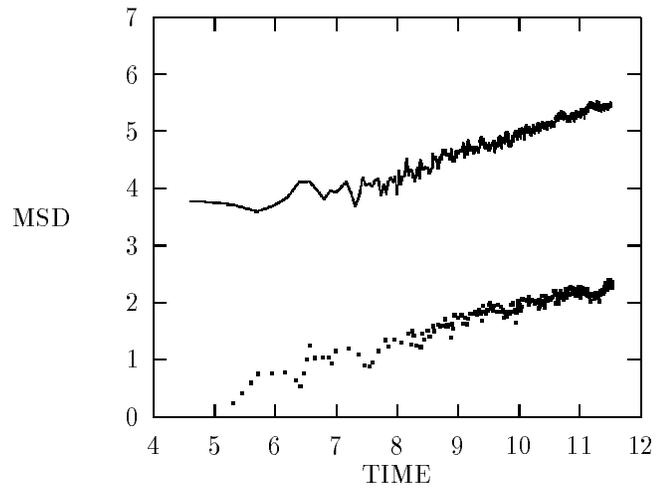
\begin{figure}
\caption{log-log plot of MSD as a function of time $t$ of a perfect
nonlinear system. (a) The solid curve corresponds to $E = 0.5$ and
$\chi = 3.5$. (b) The dotted curve corresponds to $E = 5$ and $\chi = 11$.}
\label{msd}
\end{figure}

We now discuss the behavior of quasiparticles in
 field-induced perfect nonlinear chain. The equation of motion is same
 as Eq. \ref{sch} with $\chi_m = \chi$ and
 $\epsilon_m = 0$ for all $m$.
The strength of all the sites are altered dynamically depending on the
probability of the corresponding sites. We first study for low field
strength (e.g. $E = 0.5$). The
 time-averaged probability of a few sites as a function of $\chi$ is
 shown in Fig. \ref{plp}. For small values of
 nonlinearity the time-averaged probability of the sites under study shows
smooth behavior. We find that the dynamical localization 
gets destroyed even for small values of $\chi$. However, particle gets
localized within a few sites. For further increase
of $\chi$ fluctuations, albeit small, in time-averaged probability
of the sites under study occur. The
 fluctuation regime starts from $\chi \sim 2$. In this regime MSD shows
subdiffusive behavior for a long time with many oscillations as shown 
in Fig. \ref{msd}(a). The dynamic alteration of the site
 energies makes the localization weak. The oscillating behavior in MSD
 indicates the strong
localization due to the applied electric field.
Near the self-trapping transition we find
 particle initially gets localized at the initial excitation site for some
 time. Then it gets diffused to other sites. As we move towards the
 critical value of $\chi$ the particle gets trapped at the zeroth site
for longer and longer time. This also reflects in the study of time-averaged
probability. Two sharp falls in $\lan P_0 \ran$ occur near
 $\chi_{cr}$. Thus we get the self-trapping transition at
$\chi_{cr} \sim 5.7$. It should be noted that the value of $\chi_{cr}$
 of field-induced perfect nonlinear 
system is larger than that of single nonlinear problem (compare Fig. \ref{slp} 
and Fig. \ref{plp}). In the former case the localization is weaker
 than the later one  and consequently, to get the 
self-trapping transition more $\chi$ value is needed. However, above
 $\chi_{cr}$ nonlinearity enhances the field-induced localization.

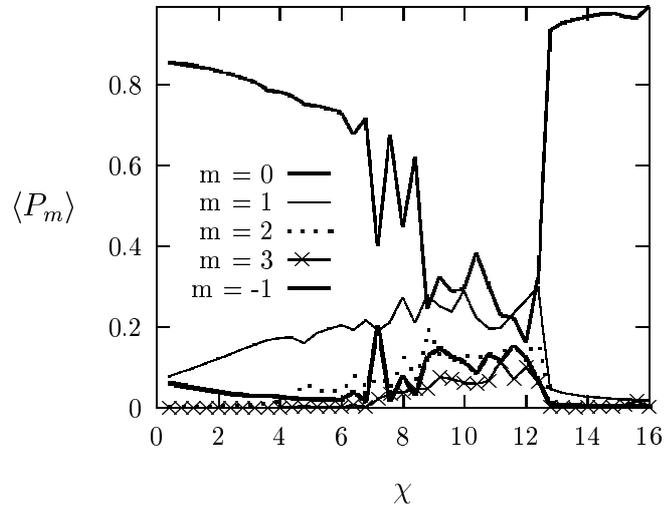
\begin{figure}
\caption{Same as Fig. \ref{plp} but $E = 5$.}
\label{php}
\end{figure}

In Fig. \ref{php} we have plotted the time-averaged probability of a few
 sites of perfect nonlinear chain for high dc field (e.g. $E =
 5$). For small values of $\chi$ they show smooth 
behavior. In this region the asymmetric particle propagation and oscillatory
 behavior in the amplitude of MSD
is also obtained. For further increase of $\chi$, 
 the fluctuations in time-averaged probabilities occur. Due to the 
alteration of the site energies
in this region ($6.9 < \chi < 12.68$) particle moves away
 from the site which is degenerate with the zeroth site at $t = 0$.
 Here also subdiffusive behavior in MSD is obtained for a long time as shown in
 Fig. \ref{msd}(b). As the energy 
difference between the sites (due to the electric field) is large, the particle
 gets localized within a few sites with maximum probability.
 Thus the value
 of MSD is much smaller than that of low field case. 
However, after self-trapping transition $(\chi_{cr}
 \sim 12.68)$ we find the
 particle gets localized at the initial excitation site and consequently
 the value of MSD is smaller than that of perfect field-induced systems. Thus 
if $\chi$ exceeds the critical value where the self-trapping transition
 occurs nonlinearity makes the localization stronger.

 In conclusion, our numerical study of the dynamics of the particle in
a perfect chain with one
 nonlinear impurity and in a perfect nonlinear chain under the action of
 low as well as high electric field indicates the existence of three
 regions of $\chi$ where the behavior is different. The values
 of $\chi$ are different for different systems and field strengths. In
 all the cases we found a sharp self-trapping transition of $\lan
 P_0\ran$ at $\chi_{cr}$. The critical value of $\chi$ depends on the
 system as well as field strength. When $\chi < \chi_{cr}$ the
 nonlinearity reduces the field-induced localization. Even we obtained
 subdiffusive behavior in field-induced perfect nonlinear
systems. However, above the critical value of $\chi$
 nonlinearity enhances the localization.

\end{document}